# High-pressure inelastic neutron scattering study of the anisotropic S = 1 spin chain [Ni(HF$_2$)(3-Clpyradine)$_4$]BF$_4$

Daniel M. Pajerowski[1,*], Andrey P. Podlesnyak[1], Jacek Herbrych[2], Jamie Manson[3]

[1]Neutron Scattering Division, Oak Ridge National Laboratory, Oak Ridge, TN 37831, USA
[2]Department of Theoretical Physics, Wrocław University of Science and Technology, 50-370 Wrocław, Poland
[3]Department of Chemistry and Biochemistry, Eastern Washington University, Cheney, WA 99004, USA

**Abstract**. [Ni(HF$_2$)(3-Clpyradine)$_4$]BF$_4$ (NBCT) is a one-dimensional (1-D), S = 1 spin chain material that shows no long range magnetic order down to thermometer temperatures of 0.1 K. Previous ambient pressure inelastic neutron scattering experiments identified NBCT to be in the large-D quantum paramagnetic phase of the D/J phase diagram, where D is the axial single-ion anisotropy and J is the intrachain superexchange. Here, we extend the previous experiments to be at a hydrostatic pressure of 0.9 GPa. By comparing to density matrix renormalization group calculations, we find D/J increases from 1.5 to 3.2 as pressure increases from 0 GPa to 0.9 GPa, which pushes the system further into the large-D phase.

*pajerowskidm@ornl.gov

## I. Introduction

Recently, there has been a correction to the trajectory for considering and discussing states of condensed matter. Namely, it is being emphasized that classification of states by ordering and symmetry is not sufficient for some phases and instead there must be a topological classification [1]. The S = 1 spin chain is one of the simpler systems that is a gateway to the panoply of proposed topological states.

One spin Hamiltonian for an anisotropic S = 1 spin chain may be written as

$$H = J \sum_i \mathbf{S}_i \cdot \mathbf{S}_{i+1} + J' \sum_{\langle i,j \rangle} \mathbf{S}_i \cdot \mathbf{S}_j + D \sum_i (S_i^z)^2 + E \sum_i \left[ (S_i^x)^2 - (S_i^y)^2 \right] \quad (1)$$

where $\mathbf{S}_i = (S_i^x, S_i^y, S_i^z)$, J > 0 is the antiferromagnetic (AFM) intrachain (super)exchange energy, J' is the interchain (super)exchange energy, the <i, j> summation is between neighboring chains, D is the single-ion axial anisotropy, and E is the single-ion rhombic anisotropy. The phase diagram for D, E, and J was calculated in reference [2] and a portion is shown in Figure 1. Further complexities may be included, such as anisotropic exchange in the so-called XXZ chains [3,4]. The bellwether S = 1 case that only considers isotropic AFM intrachain interactions has a nondegenerate gapped ground state that is called the Haldane phase [5–7]. Eventually, this Haldane phase was categorized as symmetry-protected topological phase with short-range entanglement [8]. Application of other terms in equation 1 tunes the ground state away from the Haldane phase, such as for the J' = 0 and E = 0 critical easy-plane (D/J)$_C$ = 0.96845 ratio [3] or the J' = 0 and E = 0 critical easy-axis (D/J)$_C$ = –0.32 ratio [4,9]. The large-D phase is a product state of $|S^z_i = 0\rangle$ sites and is topologically distinct from the Haldane phase. Therefore, moving across the quantum phase transition out of the Haldane phase as a function of D/J (or other parameters in equation 1) is a topological phase transition, and there is no order parameter that changes during the phase transition.

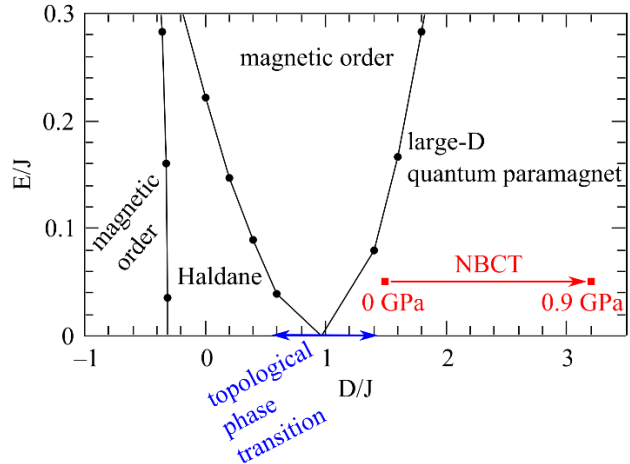

Figure 1. Anisotropic spin chain phase diagram. The theoretical boundary is from reference [2]. The point for NBCT at 0 GPa is from reference [10] and the 0.9 GPa point for NBCT is from this work.

Even before the heightened necessity of explicit topological descriptors for quantum states, the S = 1 anisotropic spin was the subject of theoretical and experimental efforts. A summary of some experimental realization of S = 1 spin chains shows that most compounds are either deep in the Haldane phase (D/J ≲ 0.25) or deep in the large-D phase (D/J ≳ 4) [11,12]. One exception being the $[Ni(HF_2)(3\text{-Clpyradine})_4]BF_4$ (NBCT) compound that bulk measurements assigned to D/J = 0.88 [13]. Inelastic neutron scattering (INS) of an isotopically doped polycrystalline NBCT compound later assigned D/J = 1.51 and E/J = 0.05 [10] based upon the spin correlations that were fit to density matrix renormalization group (DMRG) calculations. The small (<0.1 meV) gap observed for NBCT is consistent with the upper limit on the critical field ($H_C$ ≲ 35 ± 10 mT) seen in 50 mK magnetization measurements [14].

The NBCT crystal is a coordination polymer that crystallizes in the $P2_1/c$ space group [13], Figure 2. The magnetic chains are identified to be along the c-axis (c = 12.291 Å) that has F-H-F⁻ spin exchange between the formally S = 1 $Ni^{2+}$ magnetic ions, as seen in density functional theory calculations [13] and the INS [10]. There are two $Ni^{2+}$ sites within the unit cell along the chain so the AFM zone center is at a momentum of 2π/c = 0.511 Å⁻¹. These $Ni^{2+}$ sites have distorted octahedral coordination spheres with F-coordination along the c-axis chain direction and N-coordination in the plane of the octahedra. The spins have an easy-axis anisotropy (D > 0) with a preferential direction in the $N_4$ plane, as seen in the UV-visible spectroscopy [13] and the INS [10]. The pyridine rings provide a large separation between the chains (e.g. J' << J), and no magnetic diffraction peaks were found down to 0.1 K [10], as well as no long-range order signatures in specific heat or muon relaxation down to 0.3 K [13]. All experimental evidence points to NBCT being described well by J, D, and E, although there has been no formal investigation of antisymmetric or anisotropic exchange and the low symmetry of the crystal allows all three components of the antisymmetric exchange and all nine components of the exchange tensor.

This INS technique is especially diagnostic as it directly probes time and space magnetic correlations, and has been leveraged extensively for the investigation of spin chains [15]. These spin-spin correlations are observed in INS via the conjugate energy and momentum spaces. The D = E = 0 Haldane phase has triply degenerate lattice periodic spin-spin correlations $<S^\alpha S^\alpha>$ (where α = x,y,z) with the Haldane gap at the AFM zone center (often called the π-point). Numerical DMRG calculations find the Haldane phase gap at the π-point to be $\Delta = 0.41\,J$ [16]. As anisotropy is increased, this gap closes and goes to zero at the $(D/J)_C$ points described above. The introduction of axial D anisotropy breaks the degeneracy of longitudinal $<S^z S^z>$ and transverse $<S^x S^x> = <S^y S^y>$ spin-spin correlations, so two lattice periodic modes are observed in the INS. The further introduction of rhombic E anisotropy separates $<S^x S^x>$ and $<S^y S^y>$ correlations to give three distinct modes. The effects of these anisotropies on the spin-spin correlations may be captured by DMRG calculations.

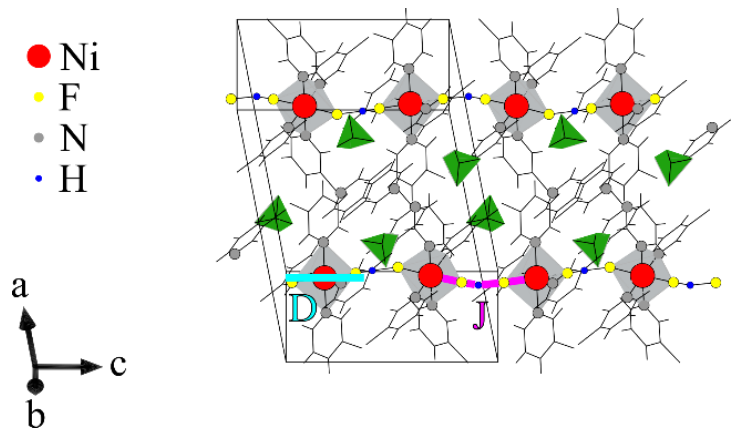

Figure 2. Crystal structure of [Ni(HF$_2$)(3-Clpyridine-D$_4$)$_4$]$^{11}$BF$_4$ (NBCT). Atoms along magnetic interaction and in nickel coordination sphere are shown, other atom representations are suppressed but all nearest neighbor bonds are illustrated. The green polyhedra are the BF$_4$ units and the rings of bonds are the (3-Clpyridine-D$_4$). Schema for the D and J parameters of equation 1 are shown. This figure is a modification of an output generated by VESTA [17].

The previous INS report on NBCT placed it at a position in the (D,E)-phase space that is precipitously close to a line separating the large-D phase assignment and a Néel phase, suggesting that external pressure could modify D, E, and J and induce a quantum phase transition. The effect of pressure for these parameters is highly specific to the material, and each term may either increase or decrease. For example, the distances and angles of the Ni-F-H-F-Ni exchange pathways of NBCT could change with pressure and modify the J parameter to either increase or decrease. Hypothetically, pressure could even tune through the Néel phase into the topologically distinct Haldane phase. The application of external pressure has previously been used with INS to show a quantum phase transition in the $S = ½$ dimer compound TlCuCl$_3$ [18]. An INS study of the Haldane spin chain Ni(C$_2$H$_8$N$_2$)$_2$NO$_2$ClO$_4$ (NENP) material showed a modification of the D/J ratio from 0.16 to 0.09 as pressure increased from ambient to 2.5 GPa, [19] but no change in phase was observed.

Here, the effect of 0.9 GPa hydrostatic pressure on the spin-spin correlations of NBCT as probed by INS is presented. The experimental 0.9 GPa spectra are quantified using DMRG calculations of the dynamical spin structure factor $S(|Q|, \hbar\omega)$ arising from equation 1 with J' = 0 to extract D, J, and E. Fits to the data show that pressure increases D/J from 1.5 at 0 GPa to 3.2 at 0.9 GPa and drives the system deeper into the large-D quantum paramagnetic phase. Additional technical details are given in Appendix A.

## II. Results and Discussion

Neutron spectra of polycrystalline NBCT were collected using a NiCrAl-alloy clamp pressure cell at P = 0.9 GPa, temperatures of T = 15 K and 0.3 K, and incident energy (Ei) values of Ei = 12.0 meV, 3.32 meV, 1.55 meV, and 1.00 meV. No magnetic signal was observed above sample energy gains of $\hbar\omega \approx 1.2$ meV. Reverse-powder-averaged, one-dimensional scattering functions were extracted from these powder data using the reported method, [20] and the one-dimensional Brillouin zone spans $Q_{1D} = [0, 2]$. The scattering from the pressure cell is greater than the scattering from more standard aluminum cans, so T = 15 K data were subtracted from T = 0.3 K data to better illustrate the spin correlations. Representative unsubtracted data are shown in Appendix B. A gapped, lattice periodic mode is visible in in the Ei = 3.32 meV spectra, Figure 3. Even at T = 15 K there is still scattering from the sample, but increasing temperature smears out the correlations as can be seen in Figure 5 (c) of reference [10], and this gives rise to an over-subtraction of the background.

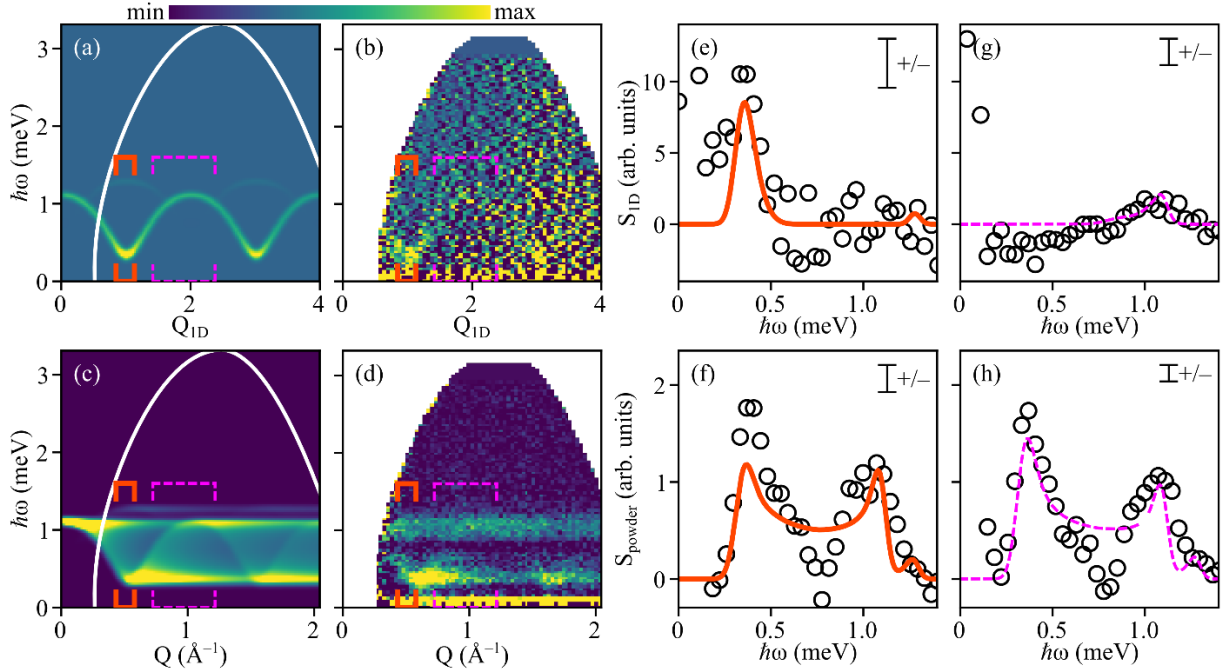

Figure 3. (note to editor: 2 column width figure) P = 0.9 GPa, Ei = 3.32 meV, subtraction of T = 15 K from T = 0.3 K neutron scattering spectra. The reverse-powder-averaged (a) model and (b) data and the powder data for (c) model and (d) data are shown. Brackets illustrate the binning regions used for the scatter plots. White lines on the model show the kinematic constraint boundary, and white regions in the data are outside of the kinematic constraint. Momentum integrations Q = [0.4335, 0.5785] Å$^{-1}$ are shown for (e) reverse-powder-averaged and (f) powder data. Momentum integrations Q = [0.7235, 1.2165] Å$^{-1}$ are shown for (g) reverse-powder-averaged and (h) powder data. The scatter plots use open circles for data and lines for the model. The uncertainty from counting statistics is given by the "+/–" bar.

To connect these experimental neutron scattering data to equation 1, the spin correlations were calculated with DMRG. A phenomenological parameterization of these spectra is possible using lattice periodic modes with functional forms motivated by linear spin wave theory, equations 3 and 4, but the resulting parameters have no built-in mapping to those in equation 1, and each $<S^\alpha S^\alpha>$ mode is independent. Instead, the approach taken here is to use DMRG calculations that have D, J, and sometimes E as inputs to connect the neutron spectra to the phase diagram in Figure 1. The DMRG spectra were themselves parameterized before comparing with experiment to reduce numerical noise, which can be especially problematic when powder-averaging the calculations, and details are in Appendix A. The lower-resolution Ei = 3.32 meV data are fit with the two intrinsic parameters D and J, and one extrinsic fit parameter that scales the overall intensity. Fitting the Ei = 3.32 meV data yields the models illustrated in Figure 3 that use D = 0.67 meV and J = 0.21 meV. The integration ranges that generate $S_{1D}$ and $S_{powder}$ as a function of ℏω were chosen to emphasize a region near the zone center (within 0.5060 Å$^{-1}$ ± 0.0725 Å$^{-1}$) and a region near the zone boundary (within 0.9700 Å$^{-1}$ ± 0.2465 Å$^{-1}$), the latter chosen to be larger due to a weaker intensity in that region. The D anisotropy splits the triply degenerate Haldane mode into two modes, $\langle S^z S^z \rangle$ and $\langle S^{x/y} S^{x/y} \rangle$, that can be clearly seen in the model plot of Figure 3 (a). Most spectral weight is associated with the $\langle S^{x/y} S^{x/y} \rangle$ band that is lower in energy at the antiferromagnetic zone center. The $\langle S^z S^z \rangle$ mode is pushed higher in energy with D and has less spectral weight than $\langle S^{x/y} S^{x/y} \rangle$. The location of the $\langle S^z S^z \rangle$ mode has an intensity consistent with the calculations, Figures 2 (f) and (h), although the intensity is close to the limit of observation. However, in the 0 GPa data [10] intensity consistent with the $\langle S^z S^z \rangle$ mode was observed, Appendix C. The bottom of the $\langle S^{x/y} S^{x/y} \rangle$ band has a gap at the antiferromagnetic zone center that increases from $\Delta_{x/y} = 0.080$ meV at 0 GPa [10] to $\Delta_{x/y} = 0.332$ meV at 0.9 GPa. In addition to the dispersive modes, there is also a mode that is flat in momentum peaked at $\hbar\omega = D$, similar to the observation at 0 GPa for NBCT. [10] The magnetic susceptibility of NBCT has a Curie tail component at low temperatures, [13] and this scattering is assigned to that paramagnetic species.

Measuring with lower incident energy reduces the flux and reduces the accessible momentum and energy transfers but provides a sharper resolution. The Ei = 1.00 meV data show the bottom of the band to be broader than the instrumental resolution, Figure 4. The solid red line in Figure 4 (c) is narrower than the spread of data at the gap. This broadening could be due to a distribution of D and J values due to pressure variations but may also have a component due to the rhombic E-term. For higher resolution data, an additional intrinsic parameter $\Delta_E$ captures the effect of the rhombic E-term by splitting the $\Delta_{x/y}$ gap of $\langle S^{x/y} S^{x/y} \rangle$ into the two modes $\langle S^x S^x \rangle$ and $\langle S^y S^y \rangle$ as $\Delta_x = \Delta_{x/y} + \Delta_E/2$ and $\Delta_y = \Delta_{x/y} - \Delta_E/2$. The splitting $\Delta_E$ may be connected to E either by DMRG or perturbation theory. Fitting these Ei = 1.00 meV data with values of D and J fixed from the Ei = 3.32 meV data yields a value of $\Delta_E = 0.07$ meV $\pm 0.01$ meV. This rhombic splitting is shown as the dashed blue line in Figure 4 (c). A similar fitting was performed in previous work of NBCT at 0 GPa to yield $\Delta_E = 0.05$ meV $\pm 0.01$ meV, [10] suggesting a small increase in E with pressure but nearly within the uncertainty estimates. These Ei = 1.00 meV data also show that there is finite intensity below the gap of the $\langle S^{x/y} S^{x/y} \rangle$ band. The source of this additional intensity could be due to multiple scattering, quasi-elastic scattering, or perhaps from a region in the sample that did not wet with the pressure medium and remained at nominally zero pressure. The statistics are not enough to make a definitive statement about the source of this intensity in the gap, although the data are suggestive that this extra scattering has an intrinsic component as it is larger at the antiferromagnetic zone center. If ≈10% of the sample were at nominally zero pressure, this extra intensity is reproduced as illustrated with the dot-dashed green line in Figure 4 (c) that could be added to the high-pressure dashed blue line to reproduce the intensities in those data. All DMRG calculations are for zero temperature so it could be that this scattering is a thermal effect, although for T = 0.3 K there is less than 0.001 % population in a 0.3 meV state above the ground state. This intensity in the gap is also present in the Ei = 1.55 meV data, Appendix D.

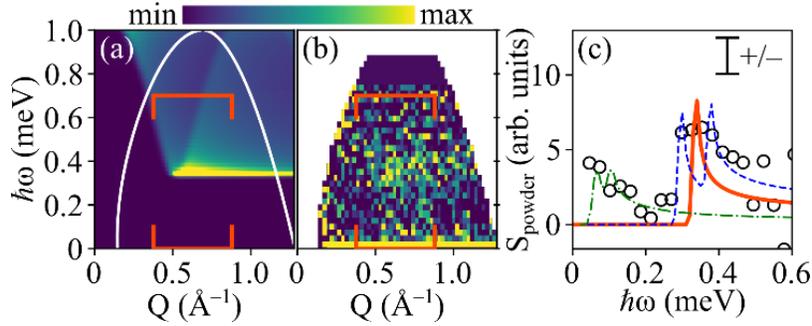

Figure 4. P = 0.9 GPa, Ei = 1.00 meV, subtraction of T = 15 K from T = 0.3 K neutron scattering spectra. The powder data for (a) model and (b) data are shown. Brackets illustrate the binning regions used for the scatter plots. White lines on the model show the kinematic constraint boundary, and white regions in the data are outside of the kinematic constraint. (c) Momentum integrations Q = [0.376, 0.880] Å$^{-1}$ are shown. The scatter plots use open circles for data. There is a thick red line for the model using D = 0.67 meV and J = 0.21 meV parameters fit from Ei = 3.32 meV ($\Delta_E = 0$). There is a dashed blue line for the model using D = 0.67 meV, J = 0.21 meV, and adding the phenomenological $\Delta_E = 0.07$ meV. There is a thin green dot-dashed line for the NBCT 0 GPa parameters D = 0.53 meV and J = 0.35 meV with $\Delta_E = 0.05$ meV. The uncertainty from counting statistics is given by the "+/−" bar.

At zero pressure, NBCT was previously identified to be precipitously close to a quantum phase transition from the large-D phase to either the Haldane phase or a Néel phase, and it was hypothesized that pressure could be used to drive the system critical [10]. These high-pressure neutron scattering data of NBCT show the gap to open up with pressure, and model fits show the system to be driven away from a quantum phase transition and instead deeper in the large-D phase, Figure 1. There is no model within the Haldane phase that accurately describes the features of these 0.9 GPa NBCT data. The effect of pressure on NBCT is illustrated with the dispersions of the $<S^zS^z>$ and $<S^{x/y}S^{x/y}>$ modes in Figure 5. The rhombic E-term (or phenomenologically $\Delta_E$) would modify these plots by splitting $<S^{x/y}S^{x/y}>$. The modifications to the spin correlations with pressure are striking, considering the modest pressure of 0.9 GPa. These pressure induced changes suggest that a material similar to NBCT may be designed that could have a pressure induced quantum phase transition. However, our result shows that for NBCT, a negative pressure either via doping or some mechanical strain is required in order to induce a quantum phase transition. Single crystal data would also be helpful to refine these powder-based models, potentially including other terms such as antisymmetric or anisotropic exchange. It will also be interesting to see how these INS determined parameters for NBCT can reproduce bulk measurements such as magnetization and specific heat.

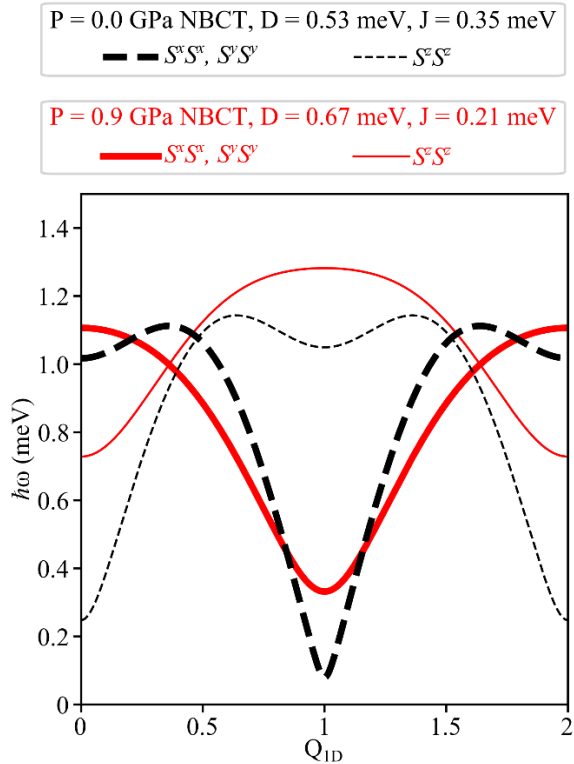

Figure 5. Pressure induced changes to model dispersion along chain.

### III. Conclusions

In conclusion, drastic changes in the spin-spin correlations of the anisotropic S = 1 spin chain [Ni(HF$_2$)(3-Clpyradine)$_4$]BF$_4$ under applied hydrostatic pressure of 0.9 GPa were observed using inelastic neutron scattering. The anisotropy increases with pressure and the exchange decreases with pressure, driving from D/J = 1.5 to D/J = 3.2 and pushing the system deeper into the large-D, with no substantial changes to E. This finding shows the large tunability of this class of coordination polymers and suggests a continued avenue of investigation for a pressure induced topological phase transition.

### ACKNOWLEDGEMENTS

Daniel M. Pajerowski and Andrey P. Podlesnyak are supported through the Scientific User Facilities Division of the Department of Energy (DOE) Office of Science, sponsored by the Basic Energy Science (BES) Program, DOE Office of Science. Jacek Herbrych acknowledges grant support by the Polish National Agency for Academic Exchange (NAWA) under Contract No. PPN/PPO/2018/1/00035. We acknowledge Mark W. Meisel for discussions and for sharing unpublished high pressure data and Gonzalo Alvarez for developing the DMRG++ code [21]. This research used resources at the Spallation Neutron Source, a DOE Office of Science User Facility operated by the Oak Ridge National Laboratory. This manuscript has been authored by UT-Battelle, LLC under Contract No. DE-AC05-00OR22725 with the U.S. Department of Energy. The United States Government retains and the publisher, by accepting the article for publication, acknowledges that the United States Government retains a non-exclusive, paid-up, irrevocable, world-wide license to publish or reproduce the published form of this manuscript, or allow others to do so, for United States Government purposes. The Department of Energy will provide public access to these results of federally sponsored research in accordance with the DOE Public Access Plan (http://energy.gov/downloads/doe-public-access-plan).

**APPENDIX A: Technical details**

A ≈0.1 gram portion of the same isotopically enriched polycrystalline samples as for the ambient pressure experiment [10] was mounted in the pressure cell and wet with Fluorinert[TM] as a pressure medium. The pressure was determined at room temperature with Raman fluorescence, and this setup has shown no significant changes of pressure between ambient temperature and low temperature [22]. Cryogenic temperatures were achieved with a wet $^3$He cryostat for the T = 0.3 K and T = 15 K data. The time-of-flight spectrometer at the SNS BL-5 (CNCS) was used in high-flux mode. [23] Experimental energy resolutions are from MANTID, and at $\hbar\omega = 0$ the full-width-half-max resolution 0.02 meV, 0.04 meV, and 0.11 meV for 1 meV, 1.55 meV, and 3.32 meV incident energies, respectively. [24] The momentum resolution is a gaussian from fitting a Bragg peak for each incident energy condition. The data were normalized to the proton current on target during collection. The detectors were normalized using a vanadium standard measurement that is defined to have an average intensity per pixel of 1 scaled unit. Intensities are multiplied by the ratio of incident and final momentum to result in numbers that are proportional to a correlation function. All numerical optimizations used the libraries of SciPy. [25]

The many-body ground state of the system is studied via the density matrix renormalization group (DMRG) method within the single-center site approach. [26–28] The dynamical correlation functions were calculated with the dynamical-DMRG method, [29,30] evaluated directly in terms of frequency via the Krylov decomposition. [31] The zero-temperature dynamical spin structure factor $S(q_{1D}, \omega)$ is evaluated as the resolvent

$$S(q_{1D}, \omega) = -\frac{1}{L\pi} \sum_{\ell} e^{i\ell q} \text{Im} \left\langle \text{gs} \middle| S_\ell \frac{1}{\omega^+ - H + \epsilon_{\text{gs}}} S_{L/2} \middle| \text{gs} \right\rangle \quad (2)$$

with $\omega = \omega + i\eta$, and |gs> ($\epsilon_{\text{gs}}$) as the ground-state wave-vector (energy). Calculations were performed for D/J = [3.0, 3.5, 4.0] using L = 80 sites on a chain with open boundary conditions. Throughout the DMRG procedure, M ≈ 1200 states are kept and ≈ 20 full sweeps are performed in the finite-size algorithm, maintaining the truncation error below $10^{-7}$. We have chosen $\delta\omega$ = 0.02 as the frequency resolution with broadening $\eta = 2\omega$.

For de-noising, these DMRG correlations were fit to the same phenomenological relationships used in the ambient pressure study of NBCT

$$I(q_{1D}, \hbar\omega) = (1 - \cos(q_{1D}))\left(I_0 + \frac{I_{-1}}{\hbar\omega}\right) \quad (3)$$

$$\frac{E(q_{1D})}{J} = \sqrt{A\cos(\frac{q_{1D}}{2})^2 + v^2 \sin^2(q_{1D}) + \Delta^2} \quad (4)$$

where the dispersion relationship parameterization is inspired by linear spin wave theory. [32] The resulting parameters were then interpolated with cubic splines to give smooth functions of A, v, $\Delta$, $I_0$, and $I_{-1}$ for the $<S^x S^x>$, $<S^y S^y>$, and $<S^z S^z>$ correlation functions. The $<S^{x/y} S^{x/y}>$ correlations were further split by letting $\Delta_{x/y} \rightarrow \Delta_{x/y} \pm \Delta_E/2$. The validity of the $\Delta_E$ parameterization to capture the resulting correlation was checked with DMRG and found to be satisfactory. To compare DMRG with experiment, a Bose factor was included to modify the intensity as a function of $\hbar\omega$. The $Ni^{2+}$ magnetic form factor was included to modify the intensity as a function of momentum transfer. [33]

**APPENDIX B: unsubtracted data with pressure cell contribution**

These high-pressure measurements are experimentally more challenging due to the smaller sample volumes and larger scattering from the containment. Neutron scattering events associated with the pressure cell and with multiple scattering from the sample position to the surrounding equipment are visible in the unsubtracted data for Ei = 3.32 meV, Figure 6. Unlike the signal from the sample, these extrinsic scattering features are essentially independent of temperature for the ranges investigated. Even on top of the large background contributions, the essential features of the data visible in the one-dimensional integrations, Figure 6 (c) and (d). There is the intensity from the bottom of the dominant mode, the intensity at the top of the mode, and the flat mode we associate with single-ion-like excitations.

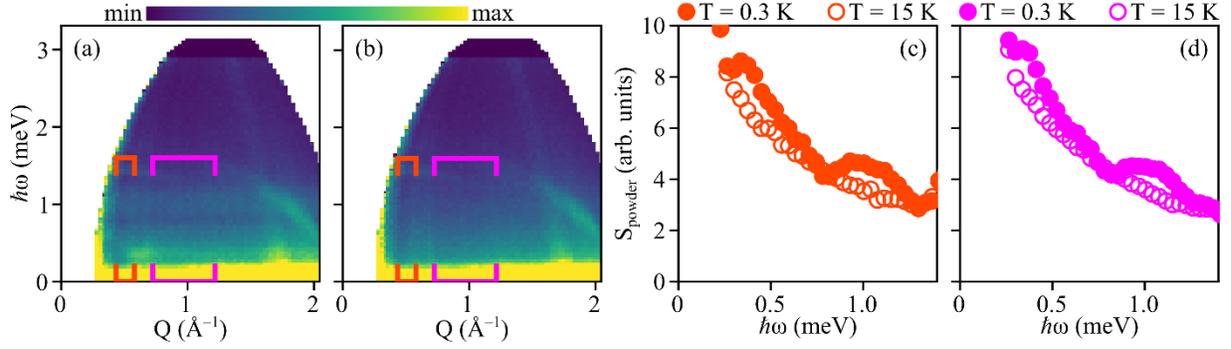

Figure 6. (note to editor: 2 column width figure) P = 0.9 GPa, Ei = 3.32 meV, unsubtracted data. (a) T = 0.3 K and (b) T = 15 K neutron scattering spectra are shown. Momentum integrations of (c) Q = [0.4335, 0.5785] Å$^{-1}$ and (d) Q = [0.7235, 1.2165] Å$^{-1}$.

## APPENDIX C: $\langle S^z S^z \rangle$ mode in 0 GPa versus 0.9 GPa

A combination of effects make observation of the $\langle S^z S^z \rangle$ mode for the high-pressure 0.9 GPa difficult. The available sample volume of the pressure cell is less than a standard can, and the extra material required to achieve high pressures decreases the transmission and increases the background scattering. Conversely, the reverse-powder-averaged 0 GPa NBCT data [10] do show scattering consistent with an $\langle S^z S^z \rangle$ mode, Figure 7.

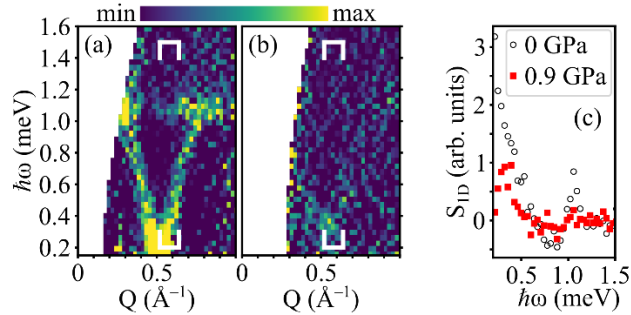

Figure 7. Comparison of Ei = 3.32 meV data showing $\langle S^z S^z \rangle$ contributions. These (a) 0 GPa and (b) 0.9 GPa data are subtractions of low-temperature minus high-temperature data visualized as a reverse-powder-average. Brackets illustrate the binning regions, Q = [0.52, 0.64] Å$^{-1}$, used for (c) the scatter plot.

## APPENDIX D: Ei = 1.55 meV

Data were also collected for Ei = 1.55 meV, Figure 8. These data are consistent with and further support the findings in the main body for Ei = 3.32 meV and Ei = 1.00 meV.

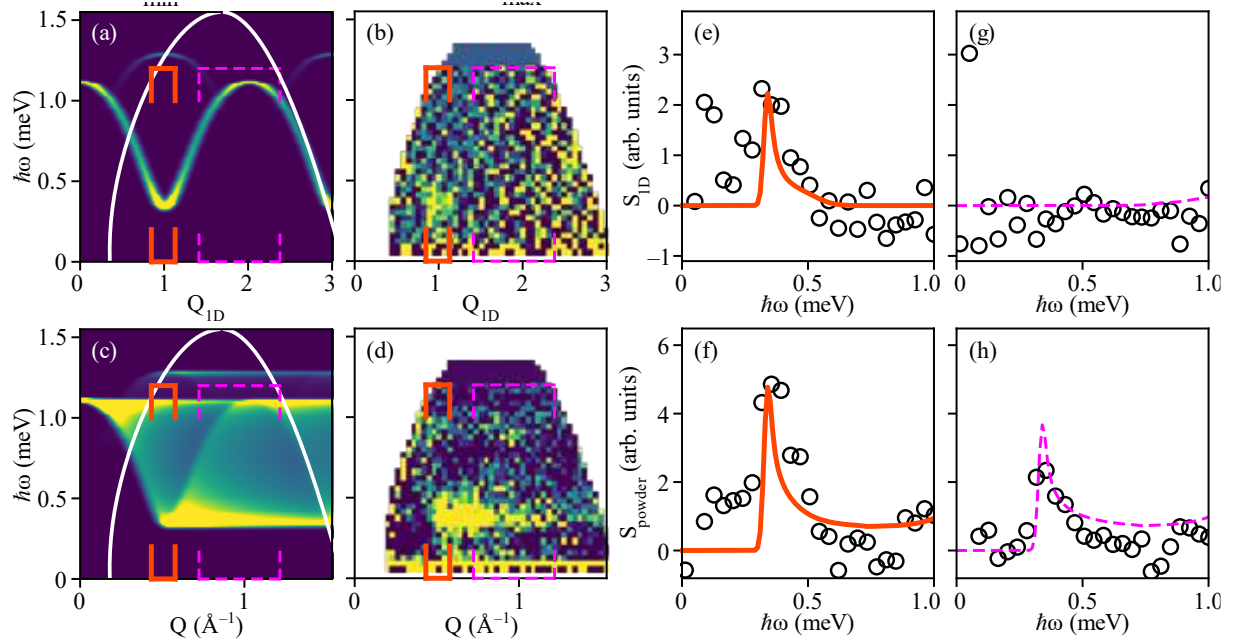

Figure 8. (note to editor: 2 column width figure) P = 0.9 GPa, Ei = 1.55 meV, subtraction of T = 15 K from T = 0.3 K neutron scattering spectra. The reverse-powder-averaged (a) model and (b) data and the powder data for (c) model and (d) data are shown. Brackets illustrate the binning regions used for the scatter plots. White lines on the model show the kinematic constraint boundary, and white regions in the data are outside of the kinematic constraint. Momentum integrations Q = [0.468, 0.628] Å$^{-1}$ are shown for (e) reverse-powder-averaged and (f) powder data. Momentum integrations Q = [0.788, 1.332] Å$^{-1}$ are shown for (g) reverse-powder-averaged and (h) powder data. The scatter plots use open circles for data and lines for the model.